\def\lesssim{{_ <\atop{^\sim}}}
\def\grtsim{{_ >\atop{^\sim}}}

\documentstyle[11pt,newpasp]{article}

\begin{document}
   
\title{Structure, dynamics and evolution of disk galaxies in a hierarchical 
formation scenario}

\author{Claudio Firmani\altaffilmark{1} and 
Vladimir Avila-Reese}

\affil{Instituto de Astronom\'{\i}a, UNAM., A.P. 70-264
 04510 M\'exico D.F., M\'exico} 

\altaffiltext{2}{Also Osservatorio Astronomico de Brera, via E.Bianchi 46, 
I-23807 Merate, Italy}

\begin{abstract}
Using galaxy evolutionary models in a hierarchical formation
scenario, we predict the structure, dynamics and evolution of 
disk galaxies in a $\Lambda $CDM universe. We find that the
Tully-Fisher relation (TFR) in the $I$ and $H$ bands is an imprint
of the mass-velocity relation of the cosmological dark halos.
The scatter of the TFR originates mainly from the 
scatter in the dark halo structure and, to a minor extension, from 
the dispersion of the primordial spin parameter $\lambda$. 
Our models allow us to explain why low and high surface brightness 
galaxies have the same TFR. The disk gas fractions predicted agree 
with the observations. 
The disks formed within the growing halos have nearly exponential
surface brightness and flat rotation curves. Towards high 
redshifts, the zero-point of the TFR in the $H$ band increases
while in the $B-$band it slightly decreases. 
\end{abstract}


\section{The method}

 We have developed a new method to study disk galaxy formation 
and evolution. Our purpose is to model self-consistently the 
structure, dynamics, and luminous properties of a disk galaxy as 
well as its evolution. At the same time, we want to predict 
correlations and statistical distributions of the disk
galaxy population.

 In our scenario of disk galaxy formation and evolution:
\begin{itemize}

\item the disk in centrifugal equilibrium forms inside-out 
within a growing dark matter (DM) halo with a rate of gas accretion proportional 
to the rate of cosmological mass aggregation (e.g., Gunn 1982; Ryden
\& Gunn 1987; Avila-Reese \& Firmani 1997;  Avila-Reese, Firmani,
\& Hern\'andez 1998), 

\item the DM halos acquire angular momentum by cosmological
torques, 

\item the star formation (SF) in the disk is triggered by gravitational
instabilities and it self-regulates through an energy balance in the 
interstellar medium (ISM).
\end{itemize}

 We assume (i) spherical symmetry and adiabatic invariance 
during the gravitational collapse of the DM, (ii) spin parameter
$\lambda$ constant in time and with a lognormal distribution,
(iii) aggregation of baryon matter to the disk in form of gas
(no mergers), (iv) detailed angular momentum conservation and 
adiabatic invariance during the gas collapse, and (v) 
stationary self-regulated SF only in the disk. 
According to our galaxy evolutionary model (Firmani, Hern\'andez, \& 
Gallagher 1996), the disk vertical structure is sustained against 
gravity by the turbulent pressure produced by the SNe and gas 
accretion kinetic energy injection. The balance of this energy input 
with the turbulent energy dissipation {\it in the disk} self-regulates 
the SF. The efficiency of SF in this model almost does not depend 
upon mass of the system, and the disk-halo feedback is 
assumed to be negligible; this is because the disk ISM 
is a dense and very dissipative medium and gas and energy outflows are 
confined in a region close to the disk.

The properties of the models depend on four initial factors: 
total virial mass $M_v$, mass aggregation history (MAH), spin parameter 
$\lambda $, and fraction of $M_v$ that is incorporated into the disk, 
$f_d$. The MAHs are calculated with the extended Press-Schechter 
formalism (Avila-Reese et al. 1998), 
the $\lambda$'s are extracted from a lognormal distribution, and $f_d$ is fixed
to 0.05. We use a flat $\Lambda $CDM model with $\Omega_\Lambda =0.65,$
$h=0.65$, $\sigma _8=1$.

\section{Results}

In Firmani \& Avila-Reese (1999a), the above method was used
to calculate catalogs of galaxy models through Monte Carlo simulations.
The main results are:

$\bullet$ A diversity of halo density profiles were obtained, the most 
typical one being close to that suggested by Navarro, Frenk \& White
(1997). Our density profiles agree rather well with those obtained
in cosmological N-body simulations (Avila-Reese et al. 1999).  

$\bullet$ The disks present a nearly exponential stellar surface 
density distribution. 
The stellar surface density (surface brightness, SB) strongly 
depends on $\lambda $, and less on mass and MAH.

$\bullet$ For a given $f_d$, the shape of the rotation curves is more peaked for
higher SB (smaller $\lambda$), but in general, the
shapes are nearly flat for most cases. If $f_d$ is too high 
($\grtsim 0.08$), the rotation curves of
disks with small $\lambda $ ($<\lambda _{\min}\approx 0.04)$ are too
peaked, and these disks are probably unstable. If $f_d$
is too small ($\lesssim 0.03$), then low and high SB galaxies
will have very different TFRs, contrary to observations. 

$\bullet$ The rotation curve decompositions show a dominance of the 
DM component down to the very central regions for
most of the models with $f_d \approx 0.05$. This occurs because the inner
density profile of the DM halos are steep ($\rho (r) \propto
r^{-1}$). When we introduce a shallow core in our DM halos ---proportional 
to the one inferred from the rotation curves of low SB galaxies---, the
disk component of the rotation curve dominates the central regions
of high SB (normal) galaxies.

$\bullet$  The slope and zero-point of the $I-$ and $H-$band TFRs are an
imprint of the mass-velocity relation of the cosmological CDM halos.
We find that the TFR is almost independent of the assumed $f_d$, when 
the disk component in the rotation curve decomposition is non-negligible
($f_d\grtsim0.03$ for the cosmological model used here).
We find that the scatter in our TFR originates {\it mainly from the 
scatter in the DM halo structure (related to the MAHs)} and, to a minor
extension, from the dispersion of $\lambda$. The predicted scatter 
does not disagree with the observational estimates.

$\bullet$ The TFR for high and low SB models is approximately the same, and
the slope of the correlation among the residuals of the TF and 
luminosity-radius relations is small and non-monotonic, although 
the shape of the rotation curves of our models correlates 
with the SB. For a given total (star+gas) disk 
mass, as the SB decreases, the maximum rotation velocity,
$V_{\max}$, of the models decreases, {\it but}, owing to the dependence of 
the SF efficiency on the disk surface density, the stellar mass $M_s$ 
(luminosity) also decreases. This combined influence of the SB ($\lambda$) 
on $V_{\max}$ and $M_s$ produces that models of different SB fall in the 
same $M_s-V_{\max}$ relation. As the result, high and low SB models 
present a similar TFR. This also explains why the $\lambda$
contribution to the scatter in our TFR becomes so small.


$\bullet$ Our models and the observations show 
that the disk gas fraction ($f_g=M_g/(M_g+M_s)$) indeed strongly correlates with 
the SB: as SB increases, $f_g$ decreases. Moreover, our 
gas fractions agree very well with the
observational estimates given in McGaugh \& de Blok (1997). 

$\bullet$ The slopes of the $H-$ and $B-$band TFRs remain almost constant
until high redshifts. For a fixed $V_{\max}$, the stellar 
mass ($H-$band luminosity) of the models at $z\approx 1$ is smaller
than at $z=0$, while, due to luminosity evolution (SF history),
$L_B$ is slightly larger (see Firmani \& Avila-Reese 1999b). Therefore, 
the observed evolution of the TFR in the $B$ band 
should not be used as equivalent of the evolution of the stellar 
mass-velocity relation ($H-$band TFR).

\end{document}